\documentclass[%
 aps,
 superscriptaddress,
 amsmath,amssymb,
reprint,%
]{revtex4-2}

\usepackage{graphicx}
\usepackage{dcolumn}
\usepackage{bm}

\usepackage[utf8]{inputenc}
\usepackage[T1]{fontenc}
\usepackage{mathptmx}
\usepackage{etoolbox}
\usepackage{subfigure}
\usepackage{todonotes}

\newcommand\Rey{\mbox{\textit{Re}}}
\newcommand\upi{\pi}

\newcommand{\dd}[1]{\mathrm{d}#1}

\newcommand{\pder}[2]{\frac{\partial#1}{\partial#2}} 
\newcommand{\tder}[2]{\frac{\dd#1}{\dd#2}} 

\begin{document}


\title{About the enstrophy change of the Reynolds-Orr solution in the channel flow}


\author{P\'eter Tam\'as Nagy}
\email[]{pnagy@hds.bme.hu}
\affiliation{Department of Hydrodynamic Systems, Faculty of Mechanical Engineering, Budapest University of Technology and Economics, Budapest, H-1111, Hungary}


\date{\today}

\begin{abstract}
	The plane Poiseuille flow is one of the elementary flow configurations. Although its laminar-turbulent transition mechanism is investigated intensively in the last century, the significant difference in the critical Reynolds number between the experiments and theory lacks a clear explanation. In this paper, an attempt is made to reduce this gap by analysing the Reynolds-Orr equation solution. Recent literature results showed that the usage of enstrophy (the volume integral of the vorticity) instead of the kinetic energy as the norm of perturbations predicts higher Reynolds numbers in the two-dimensional case. Its usage in three dimensions is discussed in the paper. In addition, other research showed an improvement of the original Reynolds-Orr energy equation using the weighted norm in a tilted coordinate system. Here, these two methods are combined. The zero enstrophy growth constraint is applied to the classical Reynolds-Orr equation, and then the solution is further refined in the tilted coordinate system.
	The results are compared to direct numerical simulations from the literature.	
\end{abstract}


\maketitle

\section{Introduction}\label{sec:introduction}

The delay of laminar-turbulent transition in the boundary layer is a promising way to achieve significant drag reduction in streamlined bodies. The transition mechanism has multiple scenarios depending on the circumstances. In the case of high free-stream turbulence level, the proper prediction method does still not exist. Furthermore, as it was pointed out by \citet{Fransson2020}, the relevant properties of the upstream flow are not clear. 

The Reynolds-Orr equation (RO), \citep{Orr1907, Schmid2001} known as the energy method, is a candidate for handling this problem. The Reynolds number is minimized where the kinetic energy change is zero.  If the Reynolds number is below the critical, minimal one, the kinetic energy of any perturbation must decay. The critical Reynolds number for the unconditional stability limit can be determined with the solution of the variational problem. This means the flow must be stable independently from the upstream flow conditions. Unfortunately, the predicted values are overly conservative and significantly below the experimental values in most cases. The explanation is that above the critical value, certain perturbations can grow for a short time, but it decays later, and it does not necessarily lead to turbulence. 

In our previous paper \citep{Nagy2019}, an active coating on boundary layer flow was investigated by the RO equation and the asymptotic solutions of the Orr-Sommerfeld equation (OS) with the aim of drag reduction. The latter method predicts the linear stability limit. The results show a considerable difference in the critical Reynolds number, and they predict different tendencies. According to the asymptotic stability analysis with the OS equation, the flow is stabilized with the increasing proportional controller parameter, while the RO equation predicts smaller values of the critical Reynolds number. According to the OS equation, with the right choice of the parameter, the critical Reynolds number can significantly increase. On the other hand, the RO equation predicts that any streamwise movement of the wall proportional to the wall shear stress slightly destabilizes the flow. The opposite trend suggests that the coating works at a low turbulence level, but it accelerates transition at a high free-stream turbulence level. However, the predicted values in the latter case are impractical. The difference between the two methods is associated with non-normality (non-orthogonal eigenvectors) of the linear problem \citep{Orr1907} and the transient growth of disturbances discussed by \citet{Schmid2007}. 
Further research is necessary to estimate the transition more reliable at high turbulence level. 

\begin{figure}
	\begin{center}
		\subfigure[]{\includegraphics[width=6cm]{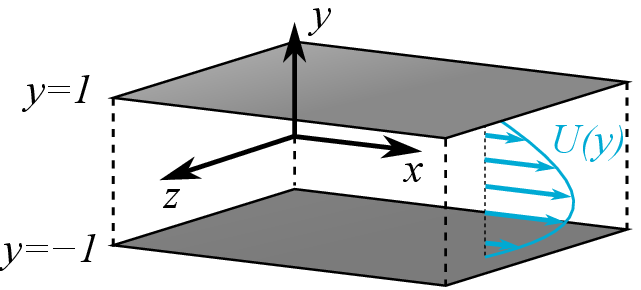} 
			\label{fig_channel}}
		\subfigure[]{\includegraphics[width=6cm]{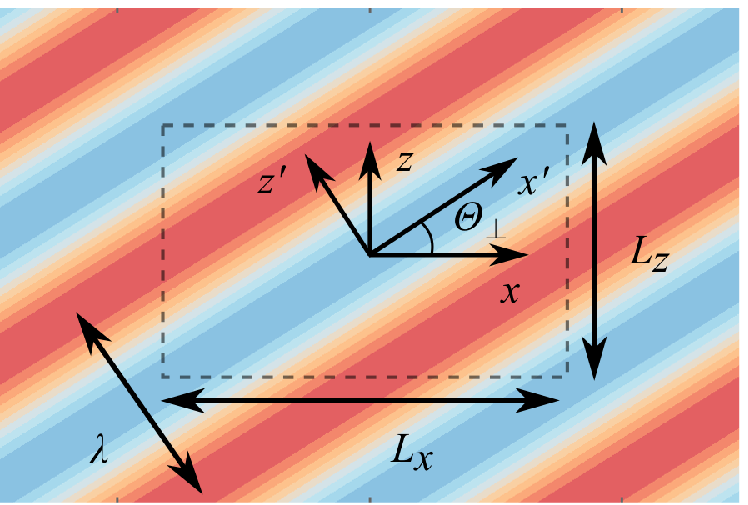} 
			\label{fig_tilted_pert}}
	\end{center}
	\caption{(a) The schematic drawing of the domain and the base flow. (b) The schematic drawing of the perturbation in the original and ($x, y, z$) the tilted ($x', y, z'$) coordinate system. The wall-normal direction ($y$) is unchanged.}
	\label{fig_tilted perturbation}
\end{figure}

Recent studies showed possible ways to improve the energy method. These methods try to predict the non-linear stability limit without prescribing the monotonic decay of the perturbations energy. These are possible ways to get practically more relevant critical Reynolds numbers. \citet{Falsaperla2019} investigated the kinetic energy of disturbances in a tilted coordinate system in the case of the plane Poiseuille (Figure \ref{fig_channel}) and Couette flow. 
They stated that the usage of the classical $L_2$ kinetic energy integral norm of perturbations, $e=||u||+||v||+||w||$ is not optimal for the non-linear stability investigation. They used a special norm of the perturbation velocity in the tilted coordinate system, where the first velocity component has a constant multiplier (weight). 
They proved mathematically that the $\frac{1}{2}\left(C||u'||+||v'||+||w'||\right)$ energy will decrease for a properly chosen $C$ parameter, if the Reynolds number is below the critical one obtained as the variation of the temporal change of $\frac{1}{2}\left(||v'||+||w'||\right)$. ($u, v,w$ are the perturbation velocity fields. The prime indicates the velocity fields in the tilted coordinate system.) This variational problem leads to the same equation as the classical Reynolds-Orr equation for a spanwise perturbation, if $\Rey_\mathrm{Orr}=\bar{\Rey}\sin\Theta_\perp$ is substituted, and the velocity fields are properly changed. $\bar{\Rey}$ is the critical Reynolds number for the weighted norm in the tilted coordinate system, while $\Rey_\mathrm{Orr}$ is the original solution. Here, the streamwise disturbance is defined as a wave that oscillates spatially in the spanwise direction, independent of the streamwise coordinate. Similarly, the spanwise disturbance does not change in the spanwise direction and oscillates spatially in the streamwise direction. The configuration can be seen in Figure \ref{fig_tilted_pert}. $x$ is the coordinate in the streamwise, $y$ in the wall-normal, $z$ in the spanwise direction. (In the cited paper, $y$ was the spanwise and $z$ the wall-normal direction. Their results are presented with the notation of this paper.) 

The authors gave the following relation between the original critical Reynolds number and the new one  
\begin{equation}\label{eq:Falsaperla}
\bar{\Rey} = \frac{\Rey_\mathrm{Orr}\left(\frac{2\upi}{\lambda\sin\Theta_\perp}\right)}{\sin\Theta_\perp},
\end{equation}
where 
$\lambda$ is the perturbation wavelength, $\Theta_\perp$ is the tilt angle when the $x'$ direction is perpendicular to the wavenumber vector. Below this Reynolds number, the previously defined norm of any single wave perturbation must decay monotonically even if the classical energy norm can increase. 
However, the two statements imply that the classical energy can grow only for a short time and must decay later. In this case, the flow is non-linearly stable, but it is not monotonically stable. From a practical point of view, the determination of the non-linear limit is more important than the strict monotonically stable limit since turbulence cannot even develop in the former case. 

At the same time, I would argue that the division with $\sin\Theta_\perp$ is not necessary in the argument of the function. In the appendix of their paper, the (A9) equation together with the continuity equation is the same as the classical Reynolds-Orr equation, after the previously mentioned substitutions. In the following steps, they used a coordinate transformation to the original coordinate system and the solution assumed in the form $v(x,y)=\tilde{v}(y)\exp(\mathrm{i}a x)$. This means that the parameter $a$ in the cited paper is the wavenumber in the original coordinate system, not in the tilted one.
In my opinion, the critical Reynolds number of weighted energy change for a tilted perturbation is 
\begin{equation}\label{eq:Falsaperlamod}
\bar{\Rey} = \frac{\Rey_\mathrm{Orr}\left(\frac{2\upi}{\lambda}\right)}{\sin\Theta_\perp}=\frac{\Rey_\mathrm{Orr}\left(\beta'\right)}{\sin\Theta_\perp},
\end{equation}
where $\beta'$ is the wavenumber in the $z'$ direction.

Although the relation (\ref{eq:Falsaperla}) or (\ref{eq:Falsaperlamod}) is similar to the Squire theorem \citep{Squire1933}, it was obtained for the non-linear energy equation instead of the linear Orr-Sommerfeld equation. However, the consequences of the two theorems are similar: the spanwise perturbations (oscillating in streamwise direction) become unstable first, since the minimum of expression (\ref{eq:Falsaperla}) is at $\Theta_\perp=\upi/2$ meaning that critical perturbation changes only $z' = -x$ direction. In this case, this equation is basically identical to the original Reynolds-Orr equation using its symmetry property. Furthermore, any streamwise perturbation (spanwise oscillating) perturbation must be stable, since $\bar{\Rey}\to\infty$ for $\Theta_\perp=0$. This statement agrees with the result of \citet{Moffat1990}, who proved that a streamwise perturbation is always stable. Falsaperla et al. generalized his theorem. This outcome seems to contradict the result of \citet{Joseph1969} who found that the critical Reynolds-number is 49.6 (using $\Rey$ definition of this paper) for a streamwise perturbation. The conflict can be resolved with the fact that the choice of classical norm used by \citet{Joseph1969} is not the best one. The weighted norm of a streamwise perturbation must decay and the flow is stable; even the classical kinetic energy of the perturbation can grow for a short time. The consequence of this is that Orr's original solution is a better estimation for the non-linear stability limit. According to the theorem of \citet{Falsaperla2019}, a spanwise perturbation is the most critical one. 

Additionally, \citet{Falsaperla2019} found good agreement with numerical and experimental results from the literature for given wavelengths and angles in Couette and Poiseuille flows. Besides, the estimated critical Reynolds number increases in both flows by roughly a factor of 2 since the streamwise disturbances, originally assumed to be more unstable, are more stable according to the new theory. However, the result is still conservative compared to experiments. Furthermore, their study cannot explain why the unstable solution at the lowest Reynolds numbers calculated with numerical simulation\citep{Paranjape2020} or experiments\citep{Prigent2003} is not purely streamwise oscillating (spanwise) perturbations. 
Since then, the authors used a similar theorem for the investigation of Bingham–Poiseuille flow \citep{Falsaperla2020},  magnetohydrodynamic flows\citep{Falsaperla2020b} and open channel flow\citep{Falsaperla2020c}.

Another interesting outcome was found regarding the critical Reynolds number when the temporal change of enstrophy instead of the weighted norm of the kinetic energy was investigated. The enstrophy is the integral of the disturbance vorticity over the whole domain. 
\citet{Fraternale2018} used this quantity in their study based on the original paper of \citet{Synge1938}. The derivation method is similar to the Reynolds-Orr equation, but the variational method applied to the temporal enstrophy change to minimize the Reynolds number. Below the minimum (critical) one, the enstrophy of the disturbances must decay, meaning that the kinetic energy must decay after a certain time. 
Unfortunately, as pointed by the authors, the temporal growth of enstrophy, contrary to the kinetic energy, is not independent of the amplitude in three dimensions. Furthermore, the variation problem is not linear but contains a quadratic term. In the derived form, their theory can be used in two dimensions only for spanwise (streamwise oscillating) perturbations where the problematic term is zero. However, the method predicts a significantly higher Reynolds number ($\Rey_\Omega=155$) than the RO equation ($\Rey=87.6$). The explanation for the difference can be interpreted as the following. Between the two Reynolds numbers, there are certain disturbance waves whose kinetic energy grows for a short time and decays later, while its enstrophy decays monotonically. Examples were shown to illustrate this in the cited paper. The explanation is similar to the result of \citet{Falsaperla2019}. There, the classical norm of the disturbance velocities can increase in a certain coordinate system for a short time, but the weighted velocity norm of the disturbance wave in a tilted coordinate system decays below the critical Reynolds number calculated with eq. (\ref{eq:Falsaperla}). The enstrophy-based stability analysis was used to investigate channel flows with blowing and suction at the walls by \citet{Lee2019}.

Many authors investigated the edge state of the channel flow numerically in the last century. Recently, \citet{Paranjape2020} carried out a thorough investigation to determine the critical Reynolds number where travelling wave perturbations can exist for a long time. The relevant literature on the problem was presented there. They used a tilted domain where the solution is spatially periodic. The velocity fields oscillate mainly in the $x'$ direction and decay in the $z'$ direction that was verified by varying the $L_{z'}$ without any effect above a certain value. They investigated the tilt angle and the streamwise length ($L_{x'}$) of the domain. For a fixed-length $L_{x'}=3.33$, the minimum Reynolds number, where the perturbation energy does not decay for a long time, was found to be 370.55 at $\Theta_\parallel = 45^\circ$. Here, the subscript of $\Theta$ indicates that in their coordinate system, the perturbation oscillates mainly in the $x'$ direction. It still depends on $z'$ variable, but it decays in that direction. At the same time, \citet{Falsaperla2019} used the tilt angle ($\Theta_\perp$) differently. The perturbation does not change in the $x'$ direction at all and spatially oscillates in $z'$. The difference between the two angles is trivial, $\upi/2$. Due to the symmetry of the problem around the $x$ axis, both angles can be defined in the range $[0, \upi/2]$ without the loss of generality. In this case, they are complementary angles.

Next, \citet{Paranjape2020} investigated the length of the domain at the fixed tilt angle $\Theta_\parallel = 45^\circ$, and the new minimum Reynolds number was found to be 367 at $L_x=3.2$. However, they cannot exclude the existence of local minimum corresponding unreported families of travelling wave solutions.

In this paper, the effect of the zero enstrophy growth constraint on the classical RO equation will be investigated in the channel flow. The hypothesis is that the edge state disturbance is close to the disturbances whose kinetic energy and enstrophy do not grow or decay. 
First, the derivation of the equation and the solution method will be introduced in Section \ref{sec:method}. Then, the equation is solved as a modal problem, and the enstrophy change is evaluated for various wavenumbers. Next, the results are shown in Section \ref{sec_results}. After that, the critical Reynolds number is improved with the formula of \citet{Falsaperla2019}. The results are compared to the numerical investigation of \citet{Paranjape2020}. Finally, concluding remarks are made in Section \ref{sec_conclusion}.

%

%
%
%
\section{The theory and the solution method} \label{sec:method}
The evolution of a perturbed flow field can be described by the following non-dimensional form of the Navier-Stokes equations
\begin{equation}\label{eq_NS_pert}
\pder{{u}_{i}}{t}=-{U}_{j}\pder{{u}_{i}}{x_j}-{u}_{j}\pder{{U}_{i}}{x_j}-{u}_{j}\pder{{u}_{i}}{x_j}-\pder{{p}}{x_i}+\frac{1}{\Rey}\pder{^2 {u}_i}{x_j^2},
\end{equation}
and the continuity equation
\begin{equation}\label{eq_cont}
\pder{{u}_{i}}{x_i}=0.
\end{equation}
${U}_{i}$ is the base flow velocity, ${u}_{i}$ is the perturbation velocity, $p$ is the pressure. $\Rey$ is the Reynolds number defined as
\begin{equation}\label{eq_Rey}
\Rey = \frac{U_0 h}{\nu},
\end{equation}
where $U_0$ is the maximum velocity at the centreline, $h$ is the half gap, $\nu$ is the kinematic viscosity. The domain is a cuboid,  $x=x_1\in[0, L_x]; y=x_2\in[-1, 1]; z=x_3\in[0, L_z]$ which can be seen in Figure \ref{fig_channel}. The domain is periodic in the streamwise, $x$ and spanwise, $z$ directions. At $y=\pm1$, no-slip wall boundary conditions hold. 
The base flow is the well-known parabolic profile:
\begin{equation}\label{eq_base_flow}
U_i = U(x_2)\delta_{i1} = (1-x_2)^2\delta_{i1},
\end{equation}
where $\delta_{ij}$ is the Kronecker delta.
The perturbation kinetic energy is
\begin{equation}\label{eq_kinetic_energy}
e = \frac{1}{2}\int_\mathcal{V} {u}_{i}^2  \mathrm{d}\mathcal{V}.
\end{equation} 
Its temporal change can be calculated by multiplying equation (\ref{eq_NS_pert}) with $u_i$ and integrating it over the whole domain. Using the Gauss divergence theorem, some terms are eliminated or rewritten, knowing that the velocity and the pressure are periodic in $x,z$ directions, and the velocity is zero at the walls. After simplification, the expression is:
\begin{equation}\label{eq_echange_pert}
\tder{e}{t} = \int_\mathcal{V} - {u}_{i}{u}_{j}\pder{{U}_{i}}{x_j} - \frac{1}{\Rey}\pder{{u}_i}{x_j}\pder{{u}_i}{x_j}
\mathrm{d}\mathcal{V}.
\end{equation}
The first term on the right-hand side is known as production and the second one is the dissipation of the kinetic energy. Minimizing the Reynold number, where the temporal change of the kinetic energy is zero, leads to a variational problem. The $i$-th component of the corresponding Euler-Lagrange equation of (\ref{eq_echange_pert}) is 
\begin{equation}\label{eq_RO_eq}
-u_j\left(\pder{{U}_{i}}{x_j}+\pder{{U}_{j}}{x_i}\right)+\frac{2}{\Rey}\pder{{}^2{u}_{i}}{x_j^2}-\pder{q}{x_i}=0,
\end{equation}
where the Lagrange multiplier, $q$ was added to the functional to prescribe divergence-free perturbations. This is the RO equation. The equations (\ref{eq_RO_eq}) and (\ref{eq_cont}) form an eigenvalue problem for the Reynolds number, and the smallest real solution is the valid one.

The enstrophy ($s$) is the volume integral of the disturbance vorticity ($\omega_i$). 
\begin{equation}\label{eq_vortex}
\omega_i = -\epsilon_{ijk}\pder{{u}_{j}}{x_k}.
\end{equation}
\begin{equation}\label{eq_enstrophy}
s = \frac{1}{2}\int_\mathcal{V} {\omega}_{i}^2  \mathrm{d}\mathcal{V},
\end{equation}
where $\epsilon_{ijk}$ is the Levi-Civita symbol. The temporal evolution of the disturbance enstrophy can be calculated as 
\begin{align}\label{eq_schange_pert}
\tder{s}{t} = \int_\mathcal{V} - {\omega}_{i}{u}_{j}\pder{{\Omega}_{i}}{x_j} + {\omega}_{i}{\Omega}_{j}\pder{{u}_{i}}{x_j}  +\nonumber\\ {\omega}_{i}{\omega}_{j}\pder{{u}_{i}}{x_j}+{\omega}_{i}{\omega}_{j}\pder{{U}_{i}}{x_j}+\frac{1}{\Rey}\omega_i\pder{^2 {\omega}_i}{x_j^2} \mathrm{d}\mathcal{V}
\end{align}
where ${\Omega}_{i}$ is the base flow vorticity. In this case, the last integrand
\begin{equation}
\int_\mathcal{V}\frac{1}{\Rey}\omega_i\pder{^2 {\omega}_i}{x_j^2} \mathrm{d}\mathcal{V}\neq \int_\mathcal{V}- \frac{1}{\Rey}\pder{{\omega}_i}{x_j}\pder{{\omega}_i}{x_j}
\mathrm{d}\mathcal{V},
\end{equation}
since the vorticity is non-zero on the walls, the term cannot be simplified with the Gauss divergence theorem. The presence of the third integrand in Eq. (\ref{eq_schange_pert}) causes that the enstrophy temporal growth rate, $\frac{1}{s}\tder{s}{t}$ depends on the amplitude of the perturbation in contrary to the kinetic energy growth rate. The third term scales with the amplitude to the power of 3  while the enstrophy and the other terms scale with the power of 2. However, this term is zero for a single periodically oscillating mode. It can be only non-zero on the periodic domain if multiple modes are present. From this point, the investigation is restricted to single wave perturbations. This assumption reduces the generality of the paper's outcome, but the results are more universal than assuming two-dimensional perturbations that were investigated recently \citet{Fraternale2018}. 
Furthermore, if the perturbation is small, the problematic term is one order of magnitude smaller than the others. The constraint is applied in the following way. The original RO equation is solved for the Reynolds number on a modal basis at various wavenumber pairs $\alpha, \beta$, and then the enstrophy change is evaluated for each mode. Since the solution of RO fulfils the zero energy change condition, the wavenumber pairs are selected where the enstrophy change is zero. The critical Reynolds number is the smallest one among these solutions. 
Since the enstrophy change constraint can be fulfilled with other multiple modes, the predicted critical Reynolds number is not a mathematically strict limit as the original method or the results of \citet{Falsaperla2019}. However, a physically reasonable assumption is that the most critical perturbation is a single wave, according to the energy theory. Here, this theory is supplemented with the enstrophy constraint. This analysis can reveal the long-living, critical travelling wave solution in the flow.

The modal solution has the form:
\begin{equation}\label{eq_Ro_Sol_form}
u_i = \hat{u}_i(y) \exp\left(\mathrm{i}(\alpha x+\beta z)\right), 
\end{equation}
\begin{equation}\label{eq_Ro_Sol_form2}
q = \hat{q}(y) \exp\left(\mathrm{i}(\alpha x+\beta z)\right),
\end{equation}
where $\alpha=2\upi/L_x$ and $\beta=2\upi/L_z$. The general eigenvalue ($\Rey$), eigenfunctions ($\hat{u}_i(y),\hat{q}_m(y) $ problem can be summarized from equations (\ref{eq_cont}) and (\ref{eq_RO_eq}) in a matrix form as 

\begin{equation}\label{eq_Operator}
\left[
\begin{array}{cccc}
2L  &  0  &  0  &  -\mathrm{i}\alpha   \\
0  &  2L  &  0  &  -D  \\
0  &  0  & 2 L  &  -\mathrm{i}\beta  \\
\mathrm{i}\alpha   &  D
& \mathrm{i}\beta  &  0  \\
\end{array}  \right]
\left[
\begin{array}{c}
\hat{u}_1\\\hat{u}_2\\\hat{u}_3\\{\hat{q}_m}
\end{array}  \right]
= \Rey 
\left[
\begin{array}{cccc}
0            & \tder{U}{y}  &  0  &  0   \\
\tder{U}{y}&  0             &  0  &  0  \\
0            &  0             &  0  &  0  \\
0   	     &  0             &  0  &  0  \\
\end{array}  \right]
\left[
\begin{array}{c}
\hat{u}_1\\\hat{u}_2\\\hat{u}_3\\{\hat{q}_m}
\end{array}  \right].
\end{equation}
after substituting Eqs. (\ref{eq_Ro_Sol_form}) and (\ref{eq_Ro_Sol_form2}) into Eqs. (\ref{eq_cont}) and (\ref{eq_RO_eq}). $L=D^2-(\alpha^2+\beta^2)$ is the Laplace operator and $D = \tder{ }{y}$ is the differential operator. The introduction of $\hat{q}_m = \hat{q}/\Rey$ is advantageous since then the numerically discretized operator on the left-hand side is not singular that reduces the number of the spurious modes appearing in the numerical calculation. 
The problem is discretized with the Chebyshev collocation method $N=60$ polynomials, similarly to the investigation of \citet{Falsaperla2019}. The wall boundary condition $\hat{u}_i(y=\pm1)=0$ is prescribed for each velocity component. The method is implemented in Matlab 2019b, and the eigenvalue problem is solved with the built-in \textit{eig} function. 

The code is verified by literature data. The minimum of the Reynolds number is determined for the most critical spanwise and streamwise perturbation. The spanwise problem is solved with series expansion by \citet{Maccreadie1931}. His minimum is $\Rey=87.63$ with this paper's notation, but he defined the Reynolds number with the gap width and the mean velocity of the base flow. The minimum with the presented technique is $\Rey=87.59368$ at $\alpha = 2.098599$. The critical Reynolds number in the case of streamwise perturbation is $\Rey=49.6035$ at $\beta=2.044$ according to \citet{Busse1969}. (The values are converted since he defined the Reynolds number with the gap width.) The minimum is $\Rey=49.603578$ at $\beta=2.043697$ with the presented technique. The values are in good agreement with literature data in both cases.

\section{The results and discussion} \label{sec_results}
\begin{figure}
	\begin{center}
	\subfigure[]{\label{fig_Re}\includegraphics[width=8cm]{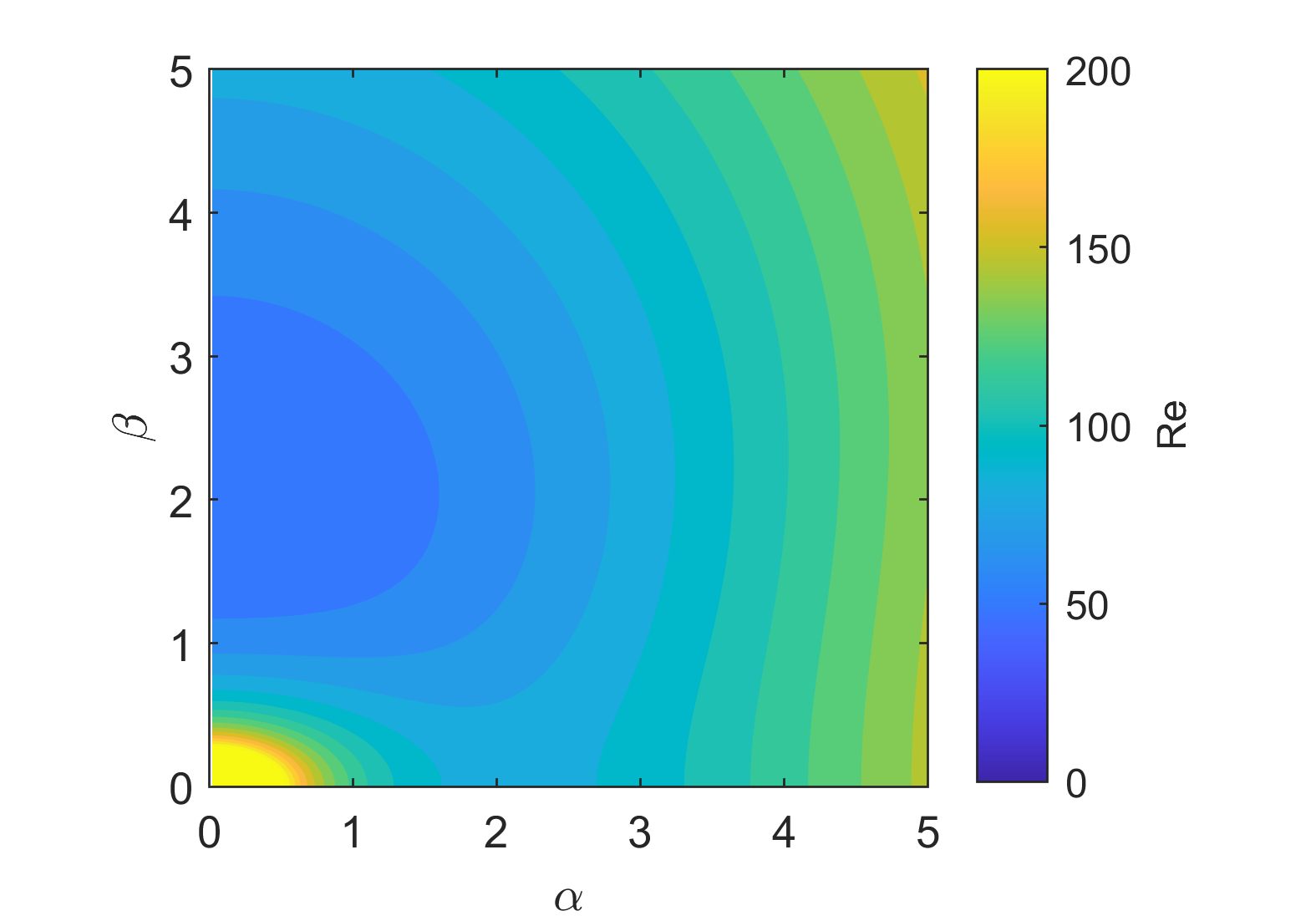}} 
	\subfigure[]{\label{fig_enstrophy_change}\includegraphics[width=8cm]{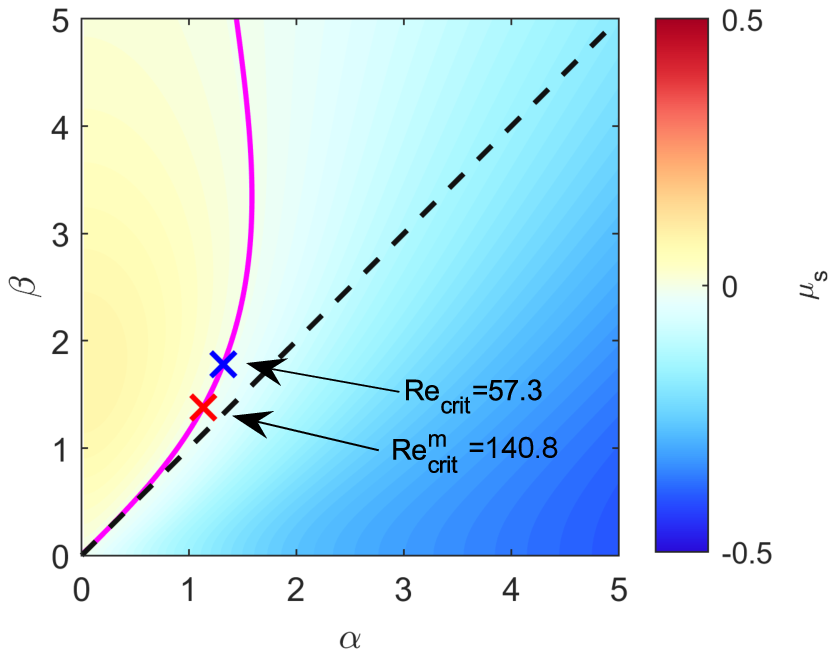}}
	\end{center}
	\caption{(a)The Reynolds number (original RO eq.) and (b) the temporal growth rate of the enstrophy as the function of streamwise ($\alpha$) and spanwise ($\beta$) wavenumbers for the most critical perturbation according to the RO equation. The magenta line represents the zero growth of enstrophy. The black dashed lines shows the perturbations where the angle between the wavenumber vector and streamwise direction is 45°. The most critical perturbation with zero enstrophy growth was found at $\alpha=1.32, \beta=1.78, \Rey_\mathrm{crit}=57.3$ according to original RO equation and at $\alpha=1.13, \beta=1.38, \Rey^m_\mathrm{crit}=140.8$ according to the method of \citet{Falsaperla2019}.}
	\label{fig_zero_growth_enstophy}
\end{figure}

The problem was solved on a fine grid of the wavenumbers. $\alpha$ is varied between 0.02 and 8 with the resolution of $\Delta\alpha=0.05$ and $\beta \in [0, 8]$ with the resolution of $\Delta\beta=0.05$. The temporal growth rate of the enstrophy 
\begin{equation}\label{ensrophy_growth}
\mu_s = \frac{1}{s}\tder{s}{t}
\end{equation}
is evaluated with equations (\ref{eq_enstrophy}) and (\ref{eq_schange_pert}). Its value is plotted in Figure \ref{fig_zero_growth_enstophy}. In the case of spanwise perturbations ($\alpha \neq 0, \beta=0$) at the bottom of the figure, the enstrophy decreases. This is the expected result since the calculated Reynolds number is much smaller than the enstrophy-based one (155) \cite{Fraternale2018}. Below the critical Reynolds number for the enstrophy change, the enstrophy must decay. However, as the tilt angle of the perturbation wave is increased, the temporal enstrophy change increases. It changes sign at $\Theta_\parallel=45^\circ$ for long waves and at higher angles for shorter waves. Furthermore, the critical Reynolds number decreases as the wavenumber vector rotates toward the spanwise direction. This means the enstrophy can grow for tilted perturbations easily, and the enstrophy-based critical Reynolds number must be lower than the energy-based one for streamwise perturbations ($\alpha=0, \beta\neq0$). The further consequence of this result is that the enstrophy-based stability analysis would predict a lower critical Reynolds number in a three-dimensional case than the kinetic energy-based one. It must be mentioned that the result of \citet{Fraternale2018} was valid only in two dimensions for spanwise perturbations and their enstrophy-based analysis predicts a much higher critical Reynolds number only in that case. The usage of an enstrophy based stability analysis is not beneficial on its own, even if the problematic term is neglected.


However, these results do not explain why not a streamwise ($\alpha = 0, \beta \neq 0$) perturbation is the most critical one. Both enstrophy and classical kinetic energy-based analysis would predict that. According to the energy theory, the critical Reynolds number is smaller in that case, and the enstrophy increases, too. The explanation can be given with the theory of \citet{Falsaperla2019} who introduced weighted norm for the kinetic energy. They showed that the critical Reynolds number with the new norm must increase as the angle of perturbations ($\Theta_\parallel$) increases. According to their theory, a spanwise perturbation should be the critical one, but the enstrophy-based analysis suggests that a streamwise perturbation is the most critical one. Furthermore, enstrophy based analysis predicts higher Reynolds number than the theory of \citet{Falsaperla2019} in the case of spanwise perturbations. The contradiction of the two results implies that the critical perturbation should be a tilted one between the streamwise and spanwise directions. This is a possible explanation for the phenomena observed in experiments\citep{Prigent2003} and numerical simulations \cite{Paranjape2020}.

In Figure \ref{fig_zero_growth_enstophy}, a single, continuous line shows the wavenumber components where both the enstrophy and kinetic energy neither grows nor decays. 
For small wavenumbers ($\alpha,\beta<1$) and large wavelengths ($\lambda>5$), the critical perturbation oscillates in the direction with an angle of 45° to the streamwise direction (black dashed curve in the figure). This finding agrees well with the literature. The systematic optimization procedure of \citet{Paranjape2020} found the most critical travelling wave solution on the domain tilted with an angle of 45°.

\begin{figure}
	\begin{center}
		\includegraphics[width=7cm]{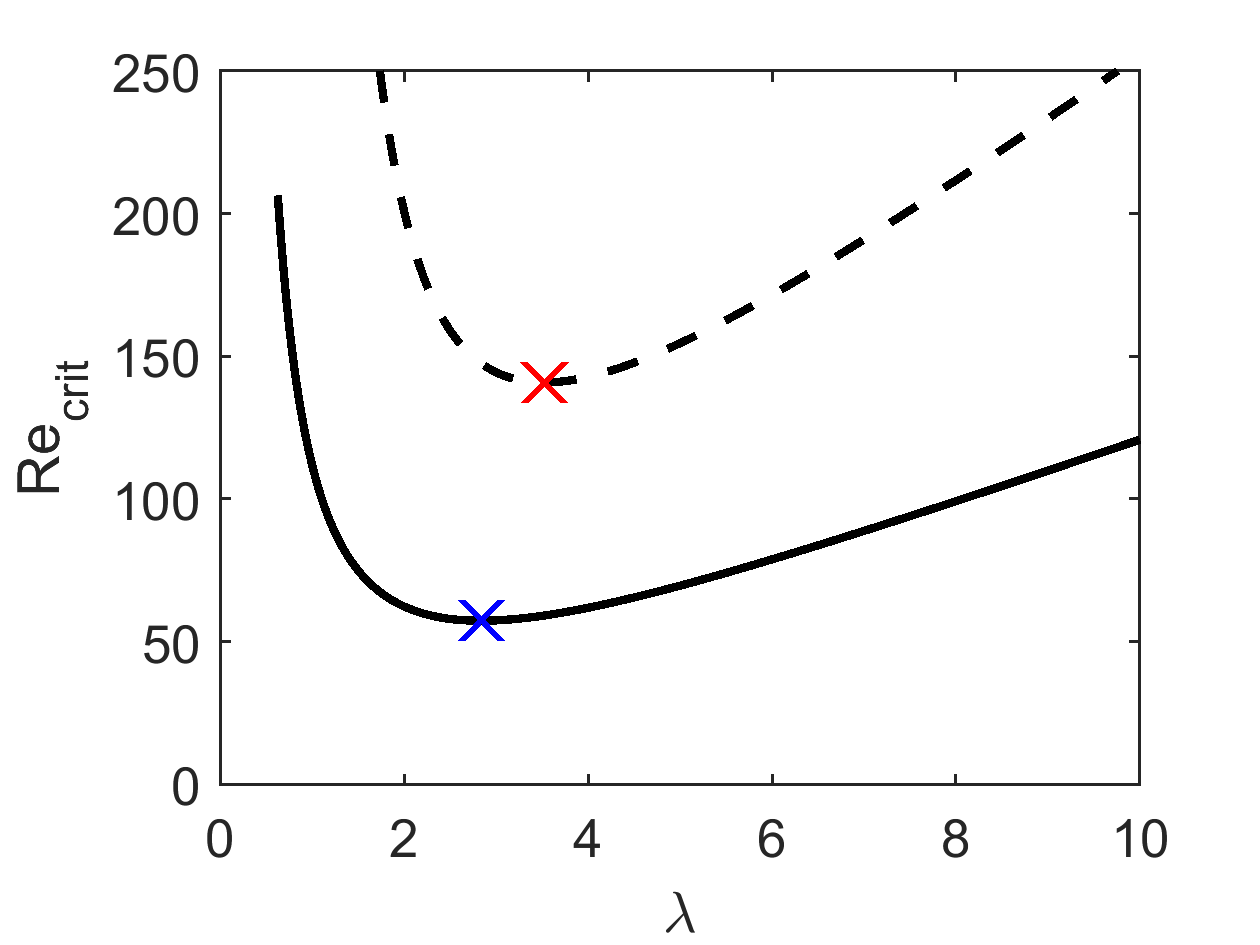}
	\end{center}
	\caption{The Reynolds number as the function of wavelength in the case of zero ensrophy growth perturbations. The continuous line is the classical solution the dashed line is the improved solution according to equation (\ref{eq:Falsaperla}). The two crosses represent the minima.}
	\label{fig_critical_Reynolds_number}
\end{figure}

In the next step, a heuristic search for the most critical perturbation is carried out. It is assumed that in the case of the critical perturbation, both the kinetic energy and the enstrophy should be in a metastable state. The wavenumber pairs such a perturbation belong to the continuous line in Figure \ref{fig_zero_growth_enstophy}. The zero enstrophy growth curve can be expressed as the function of $\beta$. In this step $\beta$ is varied between $[0.02, 10]$ with the resolution of $\Delta\beta=0.02$. The corresponding $\alpha$ value, where the enstrophy growth is zero, is determined with the built-in \textit{fminsearch} function. The initial guess comes from the previous results. For that wavenumber and tilt angle, the critical Reynolds number is estimated with the corrected formula (\ref{eq:Falsaperlamod}) of \citet{Falsaperla2019}. The wavelength and the tilt angle were determined from the wavenumber pairs. Since \citet{Falsaperla2019} assumed that the perturbations vary along the $z'$ axis, the tilt angle can be calculated as 
\begin{equation}\label{eq_theta}
\Theta_{\perp}=\left|\arctan\left({\frac{\alpha}{\beta}}\right)\right|.
\end{equation} 
The absolute value is calculated since the results are invariant to the sign of the wavenumbers, and $\Theta_{\perp}$ is defined between $[0, \upi/2]$, here. 

The critical Reynolds number calculated with the classical equation and the modified theory are plotted in Figure \ref{fig_critical_Reynolds_number}. 
Among the original theory solutions, the minimum Reynolds number of zero enstrophy growth perturbations is 57.3 at $\alpha=1.32, \beta = 1.78$, and the wavelength is $\lambda=2.83$. The angle between the streamwise direction and the wavenumber vector is $\Theta_\parallel=53.5$°. According to the modified theory of \citet{Falsaperla2019}, the critical Reynolds number increases by a factor of 2 to 140.8 at  $\alpha=1.13, \beta=1.38$. The wavelength of the most critical perturbation is  $\lambda=3.52$. The angle between the streamwise direction and the wavenumber vector is $\Theta_\parallel=50.7$°.

The smallest Reynolds number, where travelling wave solution on the tilted domain can exist, is found to be 367 at the tilt angle $\Theta_\parallel=45$° with DNS simulations by \citet{Paranjape2020}. Their angle of minima is very close to the result with the zero enstrophy growth and the improved theory. Furthermore, they varied the tilt angle in their study between 25° and 60° with the resolution of 5° first. Then, they fixed the tilt angle at 45°, and only the length of the domain was varied. The authors pointed out that their approach does not certainly predict the smallest critical Reynolds number. However, it is probably close to that value. 
The minimum Reynolds number in the numerical simulation was obtained at $L_x'=3.2$. The most energy content is associated with the perturbation wave at a wavelength equal to the size of the domain. In my analysis, the critical wavelength is 3.52 according to the improved theory that is only a 10\% longer wavelength than the DNS result. The predicted critical Reynolds number (140.8) is better than that one compared to previous theories (49.6, 87.6), but it still underestimates the value by a factor of 2.6 compared to the numerical DNS simulations (367). Furthermore, it must be mentioned that the solution of the RO equation is computationally order of magnitude less expensive than a DNS analysis. The data that supports the findings of this study are available within the supplementary material.


\section{Conclusion}\label{sec_conclusion}

The stability of channel flow is investigated. The classical Reynolds-Orr equation is solved in a modal framework, and it is improved in two ways. First, the zero enstrophy growth constrain is added to the problem. The restriction narrows down the parameter space to a single curve in the wavenumber plane. In the case of large wavelengths ($\alpha, \beta<1 \to\lambda > 5$, where the length scale is the half gap), the angle between the oscillation and the streamwise direction is 45°. Next, the theory of \citet{Falsaperla2019} is applied to these wavenumber pairs. This method predicts a significantly higher Reynolds number for tilted perturbations. The critical Reynolds number is found to be 140.8 at wavelength 3.52 with a tilt angle of 51°. The angle and the wavelength are in good agreement with numerical simulations from the literature. The predicted critical Reynolds number is still significantly smaller than one from the DNS simulations. At the same time, the new value is 3 times larger than the most critical according to classical theory (49.6) and 1.5 times larger than the theory of \citet{Falsaperla2019} (87.6). Furthermore, according to the author's best knowledge, this is the first study that explains with non-linear stability analysis why the tilted waves are the critical ones.

Applying further constraints or using other norms (weighted norms, enstrophy) may provide a more accurate estimation of the critical Reynolds number and reduce the gap between theory, simulations, and experiments in the channel transition mechanism.

\section{Acknowledgements}
	The work has been performed within the framework of the NKFI project K124939. 

\bibliographystyle{apsrev4-2}
\bibliography{biblographyV1}

\end{document}